\documentclass[final,3p]{elsarticle}

\usepackage{lineno}
\usepackage[colorlinks=true,breaklinks=true,pdftex]{hyperref}
\usepackage{adjustbox}
\modulolinenumbers[1]

\journal{GMP 2026}

\biboptions{authoryear}

\begin{document}

\begin{frontmatter}

\title{Element-Saving Hexahedral 3-Refinement Templates}


\author[a]{Hua Tong}
\author[a,b]{Yongjie Jessica Zhang\corref{cor}}
\ead{jessicaz@andrew.cmu.edu}
\cortext[cor]{Corresponding author}
\address[a]{Department of Mechanical Engineering, Carnegie Mellon University}
\address[b]{Department of Biomedical Engineering, Carnegie Mellon University}

\begin{abstract}

Conforming hexahedral (hex) meshes are widely regarded as an effective computational domain for simulation because of their nice numerical properties, yet automatically decomposing a general 3D volume into a conforming hex mesh remains a formidable challenge. Among existing approaches, methods that construct an adaptive Cartesian grid and subsequently convert it into a conforming mesh stand out for their robustness. However, topological conversion schemes require strict compatibility conditions that inevitably increase element count. State-of-the-art 2-refinement octree methods employ weakly-balanced and generalized pairing conditions to yield low element counts, but suffer from critical limitations: primal cell information is lost after dualization, and resulting dual cells often exhibit non-planar quadrilateral (quad) faces. Alternatively, 3-refinement 27-tree methods directly generate conforming hex meshes through template-based replacement, producing higher-quality elements with planar faces, but previous techniques impose far stricter conditions, severely over-refining grids by factors of ten to one hundred. This article introduces a novel 3-refinement approach using a moderately-balanced condition, slightly stronger than weakly-balanced but substantially more relaxed than prior 3-refinement requirements. The key insight is that recursively applying local refinements can isolate and reduce complex configurations to simpler cases covered by a fundamental template set. Two open-sourced variants are provided: one optimized for speed, and another trading some computational cost for marginally reduced element counts. Compared to previous 3-refinement methods, they significantly reduce final hex element counts while preserving minimum scaled Jacobian (min SJ) values and guaranteeing convex polyhedral cells; relative to 2-refinement state-of-the-art, they also achieve a lower Hausdorff ratio using slightly fewer elements.

\end{abstract}


\end{frontmatter}


\section{Introduction}
\label{sec:introduction}

Hex mesh generation is both a foundational and demanding issue within computational geometry and scientific computing. In finite element analysis (FEA) and isogeometric analysis \cite{zhang2016geometric}, hex meshes are preferred over tetrahedral alternatives because their advanced numerical properties enable improved convergence performance, reduced computational demands, and heightened accuracy for extensive physical simulation applications \cite{benzley1995comparison,cifuentes1992performance,wang2004back,wang2021comparison}. Even after long research efforts, automatically producing all-hex meshes that fit complex geometric shapes while maintaining high mesh quality still presents an unsolved challenge \cite{schneider2022large,blacker2000meeting,owen1998survey,shepherd2008hexahedral,tautges2001generation,zhang2013challenges}.

To address these challenges, several categories of techniques have emerged. Polycube map based methods construct a mapping between the input domain and an axis-aligned polycube shape to induce a regular grid, though they may encounter high distortion or failure in complex topologies \cite{gregson2011all,livesu2013polycut,guo2020cut}. The polycube methods have been developed to generate hex control nets and build volumetric spline models for isogeometric analysis \cite{zhang2012solid,wang2013trivariate,zhang2013conformal}.  Boolean operations, skeletons, and centroidal Voronoi tessellation were used to preserve geometric features \cite{liu2014volumetric,liu2015feature,hu2016centroidal,hu2017surface}. Later software packages were developed to dock with commercial software Abaqus and Ansys LS-DYNA \cite{lai2017integrating,yu2022hexgen}. Alternatively, frame-field guided parameterization methods optimize an orientation field to guide a volumetric parameterization for hex extraction \cite{nieser2011cubecover,li2012all,liu2023locally}. While these methods can produce high-quality, feature-aligned meshes, ensuring all-hex and robustness of the resulting meshes remains a significant hurdle. In contrast, grid-based methods have been distinguished as the most reliable strategy among all proposed solutions and currently serve as the only fully automatic approach that successfully transferred from academic research to industrial software \cite{MeshGems,CoreformCubit}. Typically, these algorithms start by refining a Cartesian grid adaptively until meeting specific criteria. The criteria most frequently utilized include error-sensitive functions \cite{zhang2006adaptive,zhang2010automatic,zhang2013robust,hu2013adaptive,tong2025mchex}, similarity of normals \cite{ito2009octree}, shape thickness \cite{marechal2009advances,livesu2021optimal,pitzalis2021generalized,tong2024hybridoctree_hex}, and precision of surface approximation \cite{gao2019feature,owen2017template}.

Once refinement finishes, hanging nodes throughout the grid are eliminated by replacing cells with transition templates that locally restore mesh conformity. Current methods are categorized into primal and dual methods. Primal methods directly operate on grid cells. For 3-refinement schemes, \cite{schneiders2000octree} presented the original technique, yet it struggled with concave areas, yielding excessive refinement. \cite{ito2009octree} addressed this weakness with templates for concave edges, decreasing element usage. \cite{sun2012adaptive} optimized scenarios involving isolated refinement points and geometric node locations in templates. \cite{elsheikh2014consistent} further improved the template topologies so that they can handle some concave corner cases. In comparison, 2-refinement templates, while harder to implement, can mesh both convex and concave regions with fewer elements \cite{ebeida2011isotropic,qian2012automatic,zhang2013robust,owen2017template}.

Dual method is an alternative category confined strictly to 2-refinement methods, wherein input grids are transformed so that their dual counterparts comprise purely hex elements. These methods offer a promising avenue for element usage reduction. Their main advantage relative to primal 2-refinement approaches lies in relaxing transition cell propagation to fewer than three layers. The developmental trajectory of dual methods reveals a consistent trend toward relaxing additional refinement conditions. The first technique imposed strongly-balanced conditions together with octree pairing \cite{marechal2009advances}, constraints that were preserved in two later pipelines \cite{gao2019feature,tong2024hybridoctree_hex}. Later work optimized transition templates to lower element counts \cite{hu2013adaptive,hu2016centroidal}. A substantial leap forward came through \cite{livesu2021optimal}, who introduced rotation-symmetric templates that reduced irregular edge valence and relaxed strongly-balanced condition to weakly-balanced condition while dramatically decreasing element numbers. The most element-saving method available to date has generalized octree pairing through integer linear programming \cite{pitzalis2021generalized}, further reducing element count.

The final pipeline phase projects mesh boundaries of elements within the input geometry onto the input geometry itself, concurrently maintaining mesh quality. This operation exclusively warps point-wise geometric coordinates without modifying the underlying topologies. Recent methodologies addressing this final step comprise \cite{gao2019feature,garanzha2021foldover,tong2024hybridoctree_hex,tong2025fast,maggioli2025volumetric,tong2025mchex,tong2026hexopt}.

As discussed above, a grid-based hex-meshing algorithm's element count is primarily determined by the construction of the initial grid and the removal of hanging nodes. This paper does not address the first step, nor does it make assumptions regarding the motivation behind any pre-existing refinement of the initial grid, as such decisions are application-dependent and typically derive from prior knowledge. Instead, the focus is on the second step, the refinement transformation process, whose sole objective is to ensure hex-meshability while minimizing element count increase. Specifically, this paper advances 3-refinement techniques by improving the refinement condition and introducing additional templates to mitigate over-refinement while preserving the inherent advantages of 3-refinement over 2-refinement. State-of-the-art 2-refinement methods employ weakly-balanced conditions, generalized pairing conditions, and dual meshing to produce hex meshes with minimal element counts. However, this approach suffers from some limitations: transitioning to the dual domain drops valuable information stored on primal cells, such as signed distance fields or indices of triangular faces inside primal cells, and the resulting dual elements often exhibit poor mesh quality (min SJ \(< 0.1\)) and non-planar faces, complicating practical simulation applications. In contrast, 3-refinement 27-tree methods enable direct template-based replacement of primal cells to generate conforming hex meshes with better quality (min SJ \(> 0.1\)) and planar faces. Nevertheless, existing 3-refinement approaches impose refinement conditions far more restrictive than their 2-refinement counterparts, causing excessive refinement that increases element counts by one to two orders of magnitude relative to the initial grid, thereby creating a severe bottleneck in simulation workflows.

This article introduces a novel 3-refinement approach that transforms a general adaptive grid into a conforming grid using only a moderately-balanced condition, slightly stronger than the weakly-balanced condition but substantially more relaxed than prior 3-refinement requirements. The key operation is to recursively apply local refinements to grid cells to isolate and reduce complex configurations to simpler cases, all of which are covered by a set of fundamental templates. As for implementation, two variants are proposed: a fast vertex-based version and a comprehensive edge-based version that trades computational speed for further element reduction. These two approaches reduce final hex element counts by several multiples compared to existing 3-refinement methods while preserving mesh quality and ensuring all hex elements remain convex polyhedra. Furthermore, they achieve competitive performance relative to state-of-the-art 2-refinement techniques, attaining lower Hausdorff ratios with marginally fewer elements.

The remainder of this paper is organized as follows. Section \ref{sec:refinementConditions} reviews the refinement conditions employed in prior 3-refinement methods and demonstrates how their templates yield conforming hex meshes. Section \ref{sec:fast256VertexBasedTemplates} presents the first contribution: a fast, vertex-tree-depth-based template replacement strategy. Section \ref{sec:comprehensive4096EdgeBasedTemplates} introduces the second contribution: a comprehensive but computationally more intensive edge-tree-depth-based template replacement approach with a greedy algorithm for improved element reduction. Section \ref{sec:resultsAndApplications} provides extensive experimental validation using the dataset from \cite{gao2019feature}. Finally, Section \ref{sec:conclusionAndFutureWork} concludes with a summary of contributions and directions for future research.

\section{Refinement Conditions}
\label{sec:refinementConditions}

This section begins by reviewing the conditions of 3-refinement applied to the initial 27-tree in all previous methods. Specifically, the concept of moderately-balanced condition is introduced as an effective requirement to balance transition smoothness against cell count, ensuring that all grid cells in a 27-tree can be converted into conformal hex meshes via grid cell template substitution. Subsequently, additional conditions imposed by existing 3-refinement templates are reviewed, which are essential for the practical transformation of a 27-tree into conformal hex meshes. The discussion also highlights a well-recognized limitation of these templates: their tendency to cause over-refinement of the mesh.

\begin{figure}
\centering
\includegraphics[width=\linewidth]{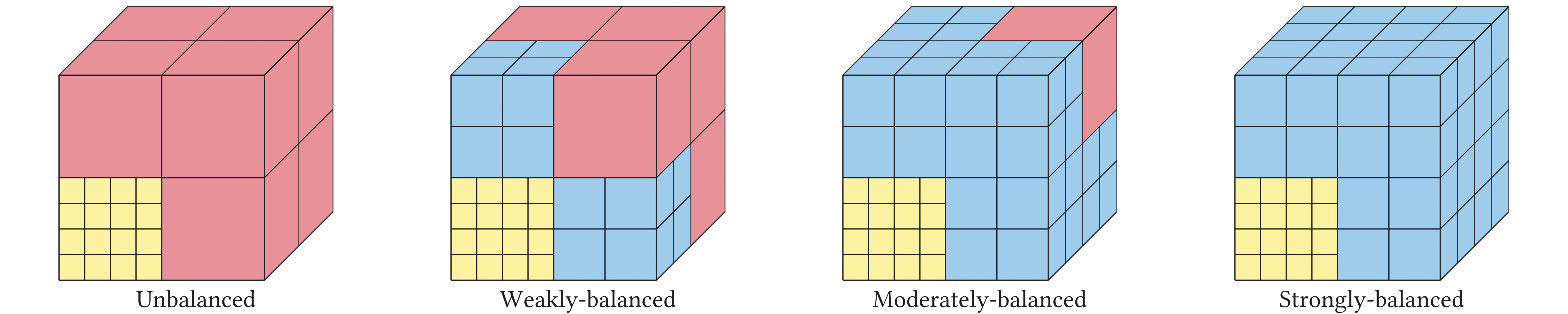}
\vspace{-8mm}
\caption{Grid cells refined under unbalanced, weakly-balanced, moderately-balanced, and strongly-balanced conditions. The total cell counts are 71 (100\%), 92 (130\%), 113 (159\%), and 120 (169\%), respectively, demonstrating increasing refinement.}
\label{fig:balancedConditions}
\end{figure}

There are four primary conditions proposed in 2-refinement for refining an initial octree: the pairing condition, generalized pairing condition, strongly-balanced condition, and weakly-balanced condition. Among them, the pairing condition and its generalized version are necessitated by a fundamental limitation of 2-refinement: it cannot independently manage the transition from a \(1 \times 1\) top face to a \(2 \times 2\) bottom face while ensuring consistent quad patterns on all four surrounding faces. The reason is that this configuration involves an odd number of quad faces, a condition proven to preclude any valid hexahedralization of the interior \cite{mitchell1996characterization}. In contrast, as will be shown in subsequent sections, 3-refinement does not suffer from this particular hex-meshability issue. Therefore, only two balanced conditions remain relevant.

The strongly-balanced condition requires that any two adjacent refined grid cells sharing a vertex, an edge, or a face have a tree level difference of at most 1. The weakly-balanced condition relaxes this, requiring a level difference of at most 1 only for cells sharing a face. This work proposes an intermediate, moderately-balanced condition, which requires that any two refined grid cells sharing an edge or a face must not differ by more than one level. Figure \ref{fig:balancedConditions} illustrates, from left to right, an unbalanced initial grid followed by the refined grids satisfying the weakly-, moderately-, and strongly-balanced conditions, with the number of elements progressively increasing. In 2-refinement, the weakly-balanced condition results in at most \(1 \to 4\) edge transitions, while the moderately- and strongly-balanced conditions result in at most \(1 \to 2\) edge transitions. All three conditions restrict at most \(1 \times 1 \to 2 \times 2\) face transitions.

Returning to 3-refinement, the three balanced conditions can be analogously defined by replacing the octree with a 27-tree. In this setting, the weakly-balanced condition allows up to \(1 \to 9\) edge transitions, while both the moderately- and strongly-balanced conditions restrict edge transitions to at most \(1 \to 3\). All three conditions restrict at most \(1 \times 1 \to 3 \times 3\) face transitions. Due to the abrupt and undesirable nature of \(1 \to 9\) edge transitions, which are generally considered too irregular for mesh generation, the weakly-balanced condition is excluded from consideration. Instead, this study employs the moderately-balanced condition, which achieves the same favorable edge transition behavior as the strongly-balanced version but with fewer constraints and lower computational cost.

Building upon the key advantage that both \(1\times1\) and \(3\times3\) faces in 3-refinement comprise an odd number of quad faces, 3-refinement can locally process transition regions independently, without introducing inconsistent quad patterns on shared interfaces between adjacent cells. This enables the adoption of a simple, direct primal approach that restores mesh conformity by replacing cells containing hanging nodes with predefined templates \cite{schneiders1996octree}. The key challenge lies in ensuring conformality across all possible configurations of cells sharing edges or faces, as well as within their interiors.

\begin{figure}
\centering
\includegraphics[width=\linewidth]{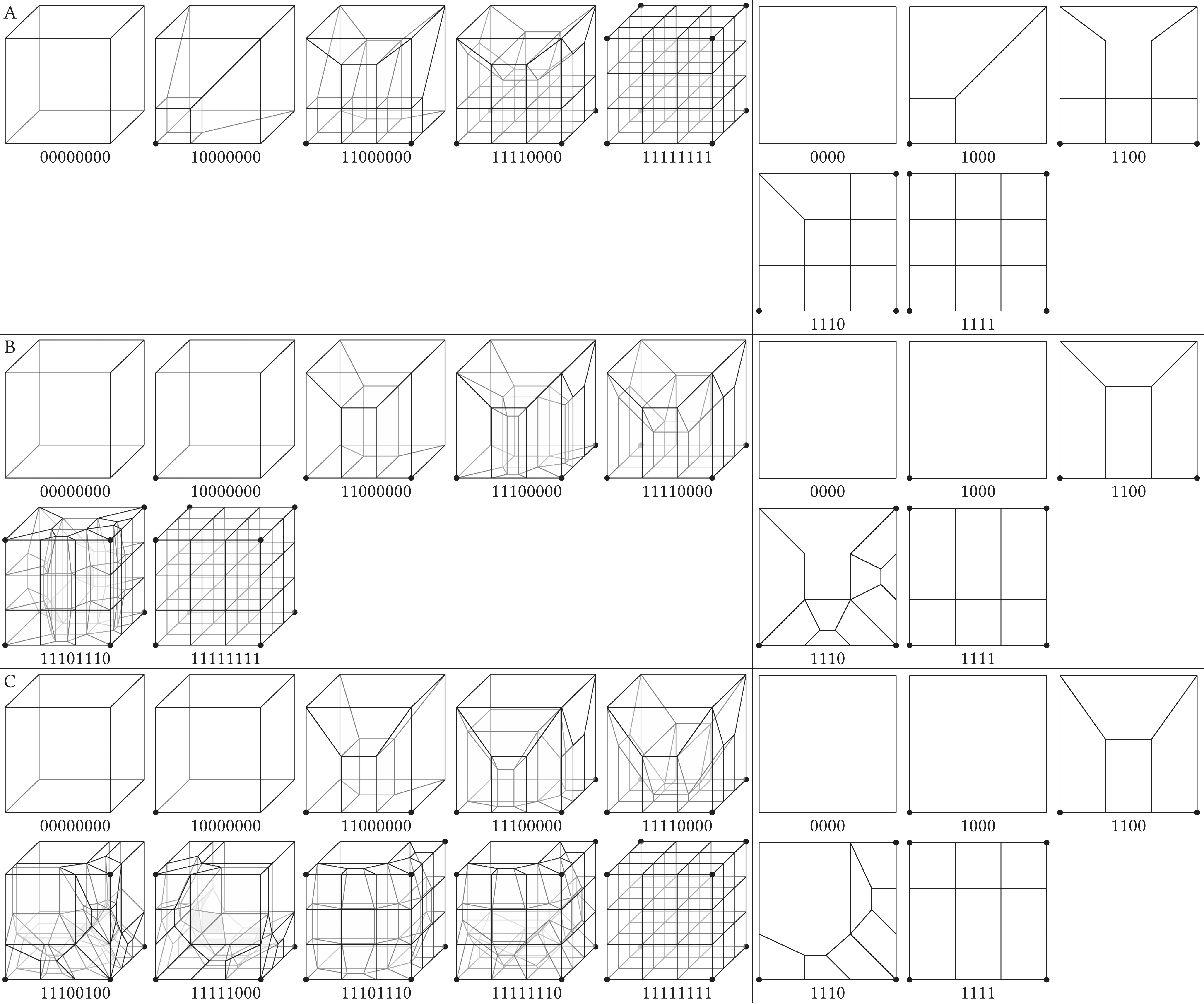}
\vspace{-8mm}
\caption{Transition templates for a grid cell (left) and a face (right) from prior work: (A) \cite{schneiders1996refining}; (B) \cite{ito2009octree}; and (C) \cite{elsheikh2014consistent}. Vertices adjacent to deeper-level cells are marked with a black dot (annotated as 1); others are annotated as 0. In all cell transition sets on the left, the cell is subdivided into hanging-node-free hex elements, with the transition across any face of the cell conforming to one of the five types in the face transition set on the right.}
\label{fig:previousTransitions}
\end{figure}

The first comprehensive set of transition patterns was introduced in \cite{schneiders1996refining}, with the face and cell transition patterns reproduced in Figure \ref{fig:previousTransitions} (A). This approach employs a vertex-based marking scheme: for each vertex of a cell, all cells sharing that vertex are examined, and if any adjacent cell has a greater tree depth, the vertex is marked with a black dot. Consequently, the leftmost and rightmost templates represent configurations with no marks and all marks, corresponding to unrefined and fully refined faces or cells, respectively, while the intermediate templates manage the transition between them. However, there are only five cell transition templates in total, which are capable of handling only convex vertex, convex edge, and face transitions. Given eight vertices each with two possible states (marked or unmarked), there exist \(2^8 = 256\) potential configurations. After eliminating duplicates through rotation and mirror symmetries, 22 distinct cases remain. A lookup table can therefore be constructed to map each case to one of the five cell transition templates, either by preserving the configuration or introducing additional black dots, as reviewed in Table I of \cite{ito2009octree}. A distinguishing feature of this approach is its unique relaxation of the strongly-balanced condition. The three intermediate transition templates are designed such that cells adjacent to black dots remain unit cubes after replacement, allowing them to be queued for refinement at the next tree depth. However, this benefit is outweighed by a severe drawback: the interface between adjacent tree depths cannot be concave when the finer region is considered interior; concave edge or corner transitions are prohibited. A common scenario in real-world applications is, for a given input geometry, where the objective is to capture surface features with fine cells while maintaining coarse cells in the interior to minimize total cell count, this approach instead forces uniform refinement of all interior cells to match the surface tree depth. Attempts to mitigate this limitation through the discovery of additional hex-meshable templates proved unsuccessful, as subsequent research efforts failed to expand cell transition templates using the same face transition templates. Consequently, despite enabling recursive refinement, the excessive and unnecessary subdivision has led to its infrequent use in modern applications.

The greatest leap in subsequent research extended the capability of 3-refinement to handling concave edges by adopting a new set of face transition and cell transition templates \cite{ito2009octree}. Its face and cell transitions are shown in Figure \ref{fig:previousTransitions} (B). Their key insight was that 1110 face could be locally decomposed into two 1100 face patterns. This decomposition enables transformation of 11100000 cell into two 11000000 cell templates via local refinement of its base face, and similarly facilitates conversion of 11101110 cell into two 11000000 cell templates and two 11110000 cell templates through local refinement of both base and top faces. However, this refinement is no longer recursive, as the quad faces attached to black dots are no longer unit squares. With only seven cell transition templates to cover the 22 possible vertex configurations, this method still relies on another refinement scheme in Table III of \cite{ito2009octree} to handle non-matching cases by adding black dots to map them onto the seven available templates. It can be readily verified that this template set is equivalent to enforcing the strongly-balanced condition while prohibiting concave corner transitions. However, in practical applications, refining cells intersected by the geometry surface inevitably introduces concave corner transitions within interior cavities. Consequently, like \cite{schneiders1996refining}, this method forces uniform refinement of all interior cells to match the surface tree depth. The sole distinction is its ability to avoid subdividing concave edges in L-shaped sweeps, yielding only modest local cell count reduction without addressing the fundamental issue of over-refinement.

Subsequent refinements upon \cite{ito2009octree} were proposed. \cite{sun2012adaptive} introduced two modifications: (1) substituting the 10000000 template in Figure \ref{fig:previousTransitions} (B) with that in Figure \ref{fig:previousTransitions} (A) when an isolated black dot yields the 10000000 pattern in all adjacent cells; and (2) offsetting the interior square of the 1110 face template in Figure \ref{fig:previousTransitions} (B) toward the upper-left corner to improve mesh density uniformity. However, these modifications are relatively minor. The first addresses a rare configuration that seldom occurs in octree/27-tree refinements as the initial adaptive refinement is usually performed on cells rather than vertices, while the second only adjusts geometric positioning without altering fundamental template topology.

Another method proposed later addresses vertex and edge irregularity while accommodating 3 concave transitions among the 22 possible vertex configurations \cite{elsheikh2014consistent}. As illustrated in Figure \ref{fig:previousTransitions} (C), this method recursively assembles non-hex-meshable cells from meshable cells through localized refinement: 11100000 cell is decomposed into two 11000000 cells via corner refinement; 11100100 cell is decomposed into three 11100000 cells; 11111000 cell is decomposed into one 11110000 and two 11100000 cells; 11101110 cell is decomposed into two 11110000 cells via edge refinement; and 11111110 cell is decomposed into three 11110000 cells. However, these transitions exhibit inherent edge inconsistencies. For example, 01 edge in the 1100 face configuration appears as a single segment, whereas the same edge in the 1110 face is split into two segments. This inconsistency stems from local refinement isolation around non-black vertices or edges connecting two non-black vertices, occurring in five patterns: 11100000, 11100100, 11111000, 11101110, and 11111110. To enforce consistency, the authors developed an algorithm that first marks these non-black vertices and edges, then processes all local refinements in adjacent grid cells before applying cell templates. Since a grid cell may border with multiple non-black vertices and edges, necessitating repeated refinements that degrade mesh quality, the algorithm directly performs 1-to-27 refinement on cells adjacent to excessive numbers of non-black vertices or edges. Additionally, to constrain refinement patterns to the ten cell transition cases, the method requires the grid to be moderately-balanced and to satisfy the pairing condition, wherein subdividing a cell requires subdividing all 27 children of its parent. While this approach successfully resolves the concave corner limitations of \cite{ito2009octree} and reduces vertex/edge irregularities, its complicated refinement criteria demand numerous iterative cycles to converge, resulting in slow computation time. Furthermore, for meshes without extensive concave corner regions, these criteria often produce more hex elements than both \cite{ito2009octree} and \cite{schneiders1996refining}.

The key challenge confronting existing methods is the trade-off between hex element count and matching of diverse face and edge transitions. This dilemma can only be resolved through more relaxed refinement conditions that minimally refine the initial grid. In this paper, two complementary template systems operating under a single moderately-balanced condition are proposed, substantially more relaxed than previous approaches. The first system (Section \ref{sec:fast256VertexBasedTemplates}) introduces vertex-based templates, enumerating all \(2^8=256\) configurations of black dot presence at a grid cell's eight vertices and providing boundary-conforming hex meshing for each scenario. The second system (Section \ref{sec:comprehensive4096EdgeBasedTemplates}) extends this concept with edge-based templates, derived from the \(2^{12}=4,096\) possible black edge configurations along a cell's twelve edges. Notably, the vertex-based templates constitute a significant extension of \cite{ito2009octree} with similarly fast computation time. The edge-based template set is a further extension; however, experimental results indicate that direct application of all 4,096 templates yields even higher element counts than the vertex-based approach. Therefore the method is augmented with a greedy algorithm that selectively subdivides previously unrefined edges, ultimately achieving modestly reduced element counts compared to the vertex-based method, with longer computation time.

\section{Vertex-Based Templates}
\label{sec:fast256VertexBasedTemplates}

\begin{figure}
\centering
\includegraphics[width=\linewidth]{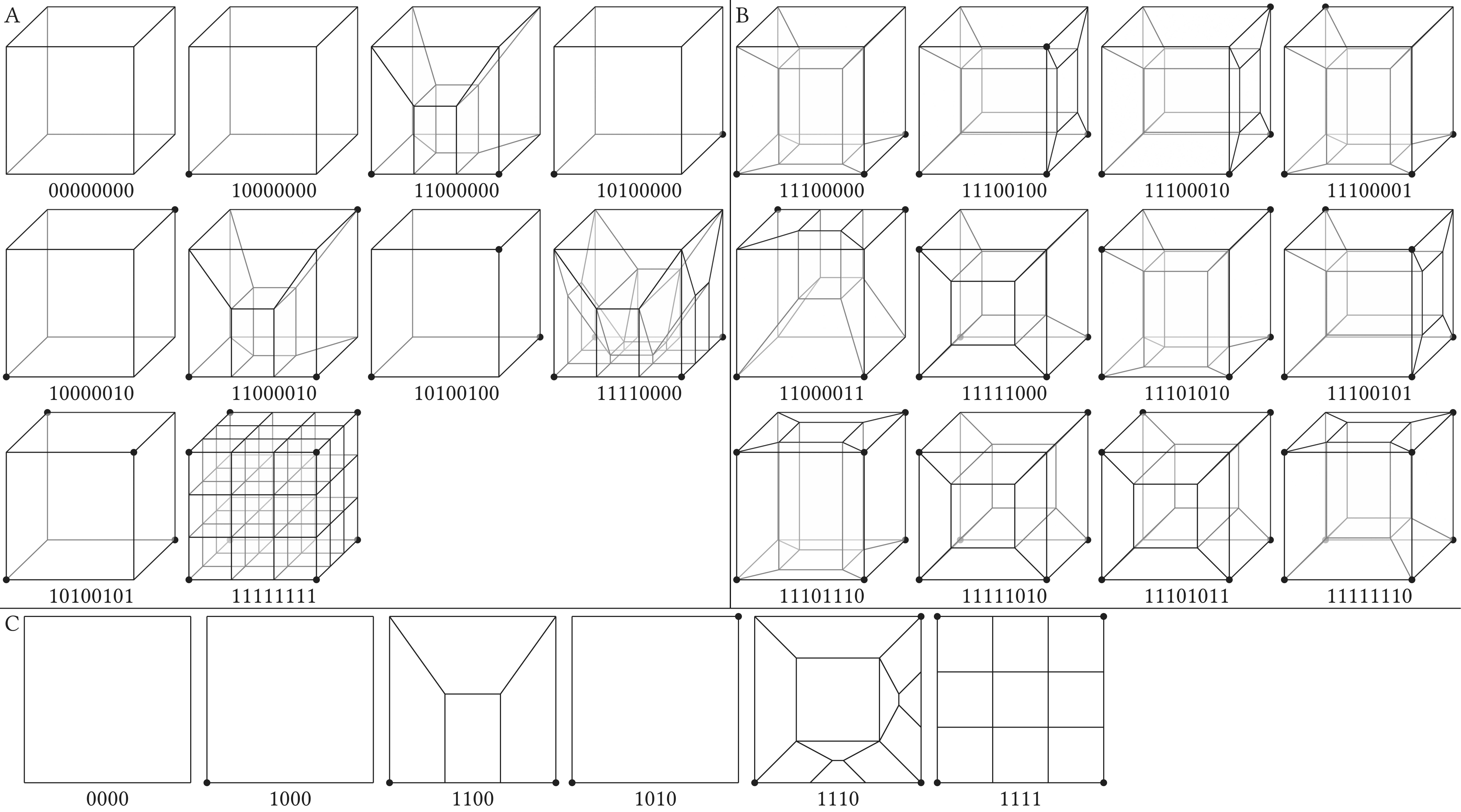}
\vspace{-8mm}
\caption{Transition templates for (A, B) a grid cell and (C) a face in vertex-based templates. Vertices adjacent to deeper-level cells are marked with a black dot (annotated as 1); others are annotated as 0. (A) The ten fundamental templates subdivide a cell into a conforming hex mesh, where each face transition is restricted to the six types in (C). (B) The twelve composite templates are obtained by applying local refinement consistent with the face transition constraints. Face conformity can be achieved by recursively applying templates to each sub-cell in a composite template.}
\label{fig:fourthTransitions}
\end{figure}

Considering the presence (1) or absence (0) of a black dot at each vertex of a cell, 22 distinct configurations are obtained after eliminating duplicates through rotational and mirror symmetries \cite{pietroni2022hex}. Figure \ref{fig:fourthTransitions} (A, B) presents the complete cell-level templates. Ten templates in Figure \ref{fig:fourthTransitions} (A) are fundamental cases derived from Figure \ref{fig:previousTransitions} (B), each readily verified to comply with the face schemes in Figure \ref{fig:fourthTransitions} (C), where edges connecting vertices of types 00 and 01 remain unchanged while type-11 edges are trisected, also ensuring edge compatibility. Twelve templates in Figure \ref{fig:fourthTransitions} (B) are composite cases, handled by extending the local refinement approach of \cite{ito2009octree} from patterns 11100000 and 11101110 to all composite configurations. Each composite cell is subdivided into five or six sub-cells, whose vertex patterns either match a fundamental template or correspond to pattern 11100000. In the former case, the sub-cell is directly replaced by its matching fundamental template; in the latter, the 11100000 template is then used, and all its sub-cells align with fundamental templates. Since all resulting sub-cells conform to fundamental templates that satisfy the face transition schemes, the final hex mesh is guaranteed to be conforming. For instance, the 11111110 composite pattern is decomposed into three sub-cells of 11110000 pattern, two sub-cells of 11100000 pattern, and one sub-cell of 00000000 pattern. Each 11100000 sub-cell is subsequently refined into two 11000000 sub-cells, two 10000000 sub-cells, and two 00000000 sub-cells, all fundamental cases, thereby completing the template substitution.

For this method, each grid edge is either preserved or uniformly trisected, requiring only a moderately-balanced condition on the input grid. Furthermore, since the algorithm operates solely through cell substitution, its time complexity is \(O(n)\) where \(n\) denotes the number of grid cells, ensuring efficient and predictable performance. In summary, the proposed templates represent a substantial extension of \cite{ito2009octree} that significantly restricts refinement propagation. While certain cells may be subdivided into more elements locally, for instance, cell 11111110 yields 27 elements when refined to 11111111 using \cite{ito2009octree} but 55 elements with vertex-based templates, the constrained propagation ultimately produces fewer global hex elements. Numerical examples in Section \ref{sec:resultsAndApplications} demonstrate that, despite occasional local increases, the vertex-based approach consistently generates fewer total elements than previous 3-refinement methods.

\section{Comprehensive Edge-Based Templates}
\label{sec:comprehensive4096EdgeBasedTemplates}

The template replacement method described in Section \ref{sec:fast256VertexBasedTemplates} adopts a vertex-based refinement criterion, consistent with prior approaches \cite{schneiders1996refining, ito2009octree, sun2012adaptive, elsheikh2014consistent}. This yields a compact set of only 22 templates, which facilitate clear presentation and straightforward implementation. However, the moderately-balanced condition requires only that edge transitions between neighboring cells be bounded by \(1\rightarrow3\) and face transitions by \(1\times1\rightarrow3\times3\). In essence, this permits each cell edge to comprise either one or three segments and each face to be either a single quad or a \(3\times3\) subdivision. Vertex-based templates, by contrast, impose stricter constraints than these conditions alone demand. As an example, consider the 1111 template face from Figure \ref{fig:fourthTransitions} (C): adjacency may occur only along the bottom and top edges with deeper-level cells. Yet under vertex-based marking, any vertex contacting a deeper-level cell must be marked as a black dot, thereby marking all four vertices and forcing the left and right edges into \(1\rightarrow3\) transitions, introducing unnecessary subdivision.

To relax the refinement, this section introduces the first edge-based template replacement method. Since a hex cell contains twelve edges, the binary state (refined or unrefined) of each yields \(2^{12}=4,096\) possible configurations. After accounting for cube symmetries via Pólya's counting theorem \cite{de1964polya}, 144 distinct edge-coloring patterns still remain, far exceeding the 22 cases in vertex-based approaches. To manage this complexity, rather than exhaustively enumerating all configurations as in Section \ref{sec:fast256VertexBasedTemplates}, special cases that leverage symmetries to reduce element counts are first identified and enumerated. Subsequently, universal templates that provide a valid decomposition for all rest edge patterns is proposed. This two-tiered method yields a comprehensive template set that substantially extends the vertex-based framework of Section \ref{sec:fast256VertexBasedTemplates}.

\begin{figure}
\centering
\includegraphics[width=\linewidth]{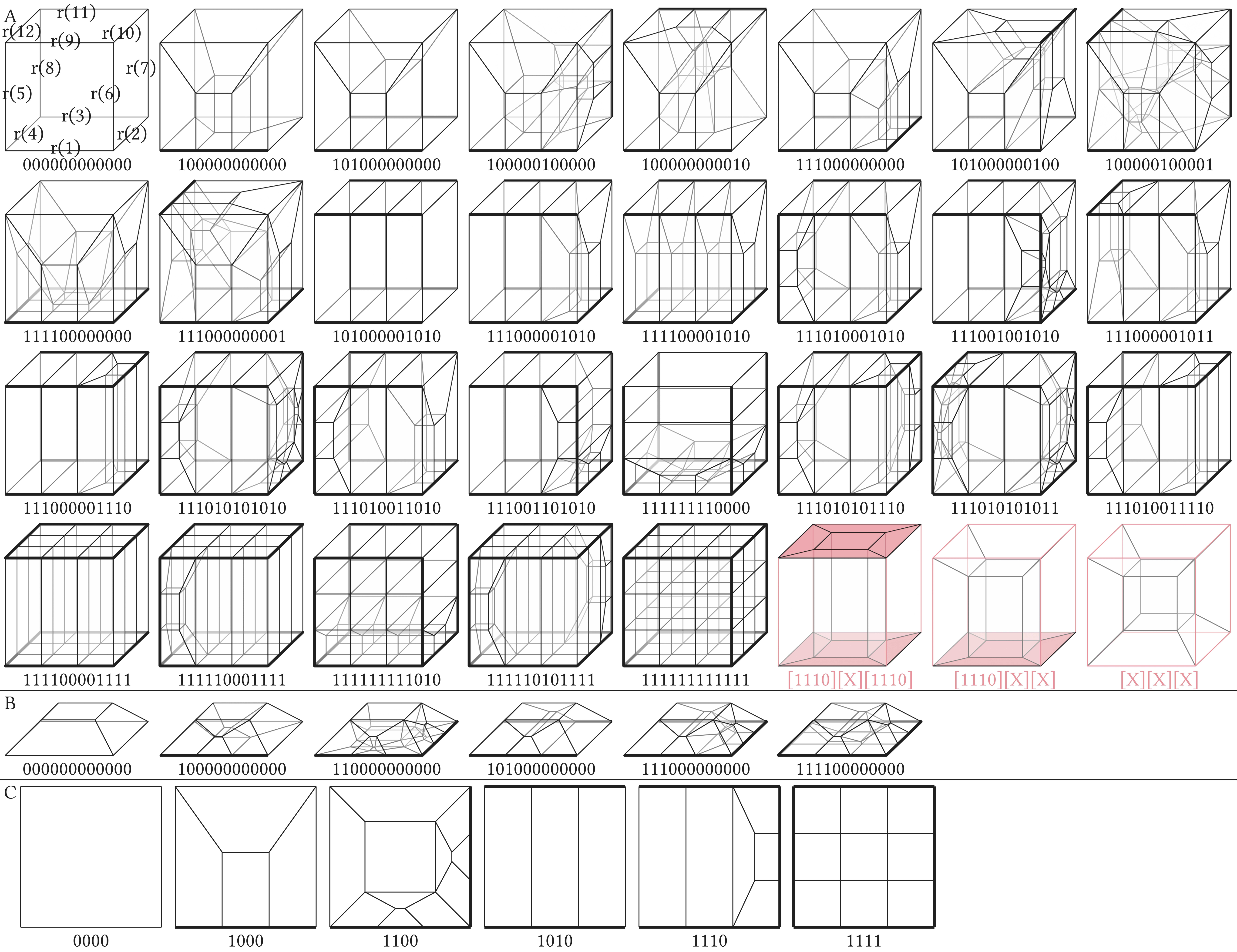}
\vspace{-8mm}
\caption{(A) Transition templates for a grid cell in edge-based templates. Edges adjacent to deeper-level cells are bolded and annotated as 1; others are annotated as 0. This is also marked in the 000000000000 cell, with 12 binary \(r(i)\) indicating if an edge is trisected and therefore bolded (1) or not (0). In all 32 cases, the cube is subdivided into hanging-node-free hex elements, with the transition across any face of the cube conforming to one of the six types in face transition schemes in (C). For the last three cases, ``[1110]'' red faces denote configurations with three 1-edges and one 0-edge, while ``X'' red edges signify arbitrary edges (may be 0 or 1). (B) Universal transition templates for all six possible sub-cell configurations in the last three [1110][X][1110], [1110][X][X], and [X][X][X] cases, where ``[X]'' represents ``XXXX''. (C) Square face transition schemes with consistent edge patterns.}
\label{fig:fifthTransitions}
\end{figure}

As illustrated in the first 000000000000 cell in Figure \ref{fig:fifthTransitions} (A), a grid cell's subdivision state can be encoded using twelve binary indicators \(r(i) \in \{0,1\}\) for \(i=1,2,\dots,12\), where ``0'' denotes an edge without trisection and ``1'' denotes an edge with trisection (bolded). The algorithm begins by examining simplified configurations leveraging symmetry. After exhaustive enumeration, the first 31 specialized cases (excluding the final [X][X][X] cell) are presented in Figure \ref{fig:fifthTransitions} (A), where ``[1110]'' represents a face with three 1-edges and one 0-edge, and ``[X]'' represents ``XXXX''. The replacement procedure, as implemented in practice, proceeds hierarchically. The edge encoding of the cell under consideration is first matched against the first 29 templates among these 31 cases, specifically excluding the [1110][X][1110] and [1110][X][X] cells. Since the encodings for these 29 templates are uniquely determined, they require no prescribed processing order; upon locating a match in the edge encoding hash table, the corresponding replacement is applied immediately. If no match is found, the algorithm next evaluates the 30th template [1110][X][1110], which identifies configurations where a pair of opposite faces each contain three edges coded as ``1'' and one edge coded as ``0''. This template isolates the four side faces into sub-cells, each of which is processed using the universal templates from Figure \ref{fig:fifthTransitions} (B) according to its side-face edge encoding. Subsequently, the 31st template [1110][X][X] is assessed, matching cells where a single face has three edges coded as ``1'' and one edge coded as ``0''; it isolates the remaining five faces into sub-cells handled via Figure \ref{fig:fifthTransitions} (B). Finally, cells matching none of the first 31 templates are processed using the 32nd template [X][X][X], which isolates all six faces into sub-cells that are handled via Figure \ref{fig:fifthTransitions} (B). The reason why the last case can cover all the cells that do not match the first 31 templates is because in the most general case, such a cell can always be partitioned via local refinement into seven sub-cells. Up to rotational symmetry, the twelve-edge subdivision encodings for these sub-cells are: \(0,0,0,0,\mathbf{0}\); \(r(1),r(2),r(3),r(4),\mathbf{0}\); \(r(11),r(10),r(9),r(12),\mathbf{0}\); \(r(9),r(6),r(1),r(5),\mathbf{0}\); \(r(3),r(7),r(11),r(8),\mathbf{0}\); \(r(4),r(8),r(12),r(5),\mathbf{0}\); and \(r(10),r(7),r(2),r(6),\mathbf{0}\), where ``\(\mathbf{0}\)'' represents that the remaining eight bits are zero. Since the last eight bits of each encoding are invariably zero, the template design is simplified, requiring only the four edge states per face. This yields the self-consistent, edge-based face template scheme shown in Figure \ref{fig:fifthTransitions} (C). Applying these six face templates to a sub-cell in the last [X][X][X] cell of Figure \ref{fig:fifthTransitions} (C) produces six conforming templates in Figure \ref{fig:fifthTransitions} (B), thereby ensuring the effectiveness of the universal scheme. However, the universal scheme is inefficient, generating substantial redundant elements. For example for the 111111111111 cell, it produces one central 000000000000 cell and six face-corresponding 111100000000 cells, totaling 79 cells, far exceeding the optimal 27-cell subdivision. This inefficiency necessitates the design of the first 31 cell-efficient templates that exploit inherent symmetries for specialized cases. Consequently, conformal hex mesh templates can be constructed for all 4,096 possible edge subdivision configurations using this comprehensive approach.

While the edge-based templates represent a substantial expansion over their vertex-based counterparts, empirical evaluation reveals that direct application often yields hex element counts comparable to, or even exceeding, those from vertex-based methods. This counter-intuitive outcome arises because the universal templates generate excessive elements, one sub-cell per refined face, without guaranteeing a net reduction; and fewer marked edges do not necessarily translate to fewer hex elements. Consequently, an optimization procedure becomes essential, one that strategically marks certain edges as subdivided to explore configurations with lower element counts. A globally optimal approach would require integer linear programming while enforcing the moderately-balanced condition, a computationally expensive approach. To balance efficiency with hex element count outcome, a local greedy algorithm is employed: first, all currently unsubdivided edges whose refinement would not violate the moderately-balanced condition are collected into set \(E\); then, each edge \(e \in E\) is evaluated sequentially, and if subdividing \(e\) does not increase the total hex element count across all affected cells, the refinement is performed. This iterative process continues until no subdivisions that reduce hex element count remain. Experimental results demonstrate that this greedy approach already yields effective reductions in hex element counts, though it is not optimal. A more advanced optimization strategy is therefore deferred as future work.

\section{Results and Applications}
\label{sec:resultsAndApplications}

To validate the vertex-based and edge-based approaches, a C++ prototype is implemented. To ensure full reproducibility, the complete source code is provided in \href{https://github.com/CMU-CBML/Element-Saving-Hexahedral-3-Refinement-Templates}{https://github.com/CMU-CBML/Element-Saving-Hexahedral-3-Refinement-Templates}. The implementation employs a traditional grid-based framework to generate an adaptively refined grid that accurately approximates the input geometry according to the shape diameter function criterion \cite{shapira2008consistent}, as adopted from \cite{pitzalis2021generalized}. The grid is then refined into conformal hex meshes using different methods. A key implementation detail is that \cite{pitzalis2021generalized} employs CGAL's shape diameter function \cite{fabri2009cgal}, which faithfully implements \cite{shapira2008consistent} by casting only inward rays. In grid-based meshing pipelines, this design may be problematic, as it neglects narrow external gaps in the input geometry, potentially causing surface merging. The implementation addresses this limitation by casting both inward and outward rays, always producing more conservative shape diameter values and more refinements near narrow external gaps. This further demonstrates the element-saving capability of the proposed approaches.

Comparisons with previous 3-refinement methods \cite{schneiders1996refining, ito2009octree, elsheikh2014consistent} are straightforward: hex element counts are directly compared from the same initial grid using the same shape diameter function parameter. The dataset comprises all 202 models from \cite{gao2019feature}, including 93 organic and 109 CAD models of varying geometric and topological complexity. Depending on the application, either the interior or exterior of the input object may be relevant, for example, structural finite element analysis primarily concerns the interior, whereas fluid mechanics simulates around the object (e.g., an aircraft or vessel). For this reason, the total number of hex elements is reported rather than only those interior to the input geometry.

\begin{figure}
\centering
\includegraphics[width=\linewidth]{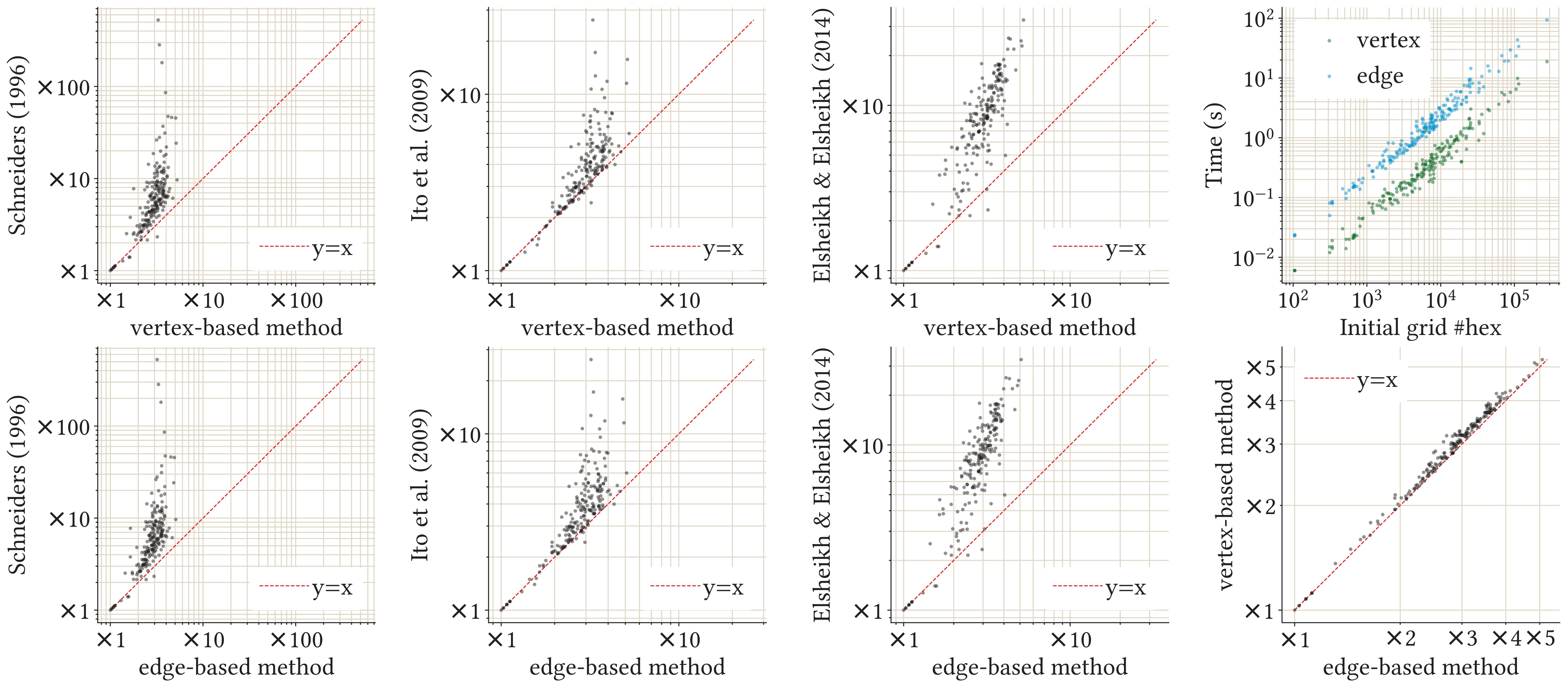}
\vspace{-8mm}
\caption{Performance comparison on 202 models in the dataset of \cite{gao2019feature}. All panels except the upper right show relative growth in hex element count versus the initial grid: the x- or y-axis represents ratios of final to initial hex elements for the vertex-based or edge-based method or previous 3-refinement methods \cite{schneiders1996refining, ito2009octree, elsheikh2014consistent}. The upper right panel displays computation time versus initial grid cell count for the proposed methods.}
\label{fig:resultComparison}
\end{figure}

Comparison with the most element-efficient 2-refinement method \cite{pitzalis2021generalized} is more complex due to different initial grids. To control variables, it is first ensured that the implementation of the shape diameter function matches theirs. For each test model, hex meshes are generated using both methods, tuning parameters such that the proposed method produces a slightly lower element count. Then exterior elements are removed using \cite{tong2024hybridoctree_hex} and the boundary is projected onto the input geometry using \cite{tong2025fast, tong2026hexopt}. Finally, the Hausdorff ratio is compared between the final mesh boundary and the input geometry; the method achieving the lower is considered to be superior. In this process, the removal of exterior elements adopts a two-step process to relax scaled Jacobian bounds to padded hex elements before projection. The deformation is driven by an optimization process coupled with a Laplace smoothing operator. Constraints relocate each boundary vertex to its closest point on the input geometry. The objective function avoids hex flips by employing a combined scaled Jacobian and Jacobian approach: the Jacobian term is specifically applied to elements with negative Jacobians. This objective function can effectively drive flipped elements back to a valid state in practice. This optimization runs for 600 seconds per mesh. Due to this pipeline introducing additional steps that are independent of the core focus of this paper, and they may not yield optimal results in complex scenarios, the Hausdorff ratio results should be interpreted with caution. However, since the contribution focuses on designing hex-meshable templates, the method can be coupled with existing robust boundary projection algorithms, potentially enhancing geometric fidelity using the same number of elements.

\begin{table}
\centering
\caption{Statistical summary of Figure \ref{fig:resultComparison}. For each comparison, the table reports three numbers: the count of models where the x-axis method achieves lower (win), equal (tie), and higher relative growth (loss) compared to the y-axis method.}
\label{tab:resultComparison}
\begin{tabular}{l|l|l|l|l}
& \cite{schneiders1996refining} & \cite{ito2009octree} & \cite{elsheikh2014consistent} & Vertex-based \\\hline
Vertex-based & 188/9/5 & 153/25/24 & 187/9/6 & 0/202/0 \\\hline
Edge-based & 188/9/5 & 177/14/11 & 188/9/5 & 185/17/0
\end{tabular}
\end{table}

\begin{figure}
\centering
\includegraphics[width=\linewidth]{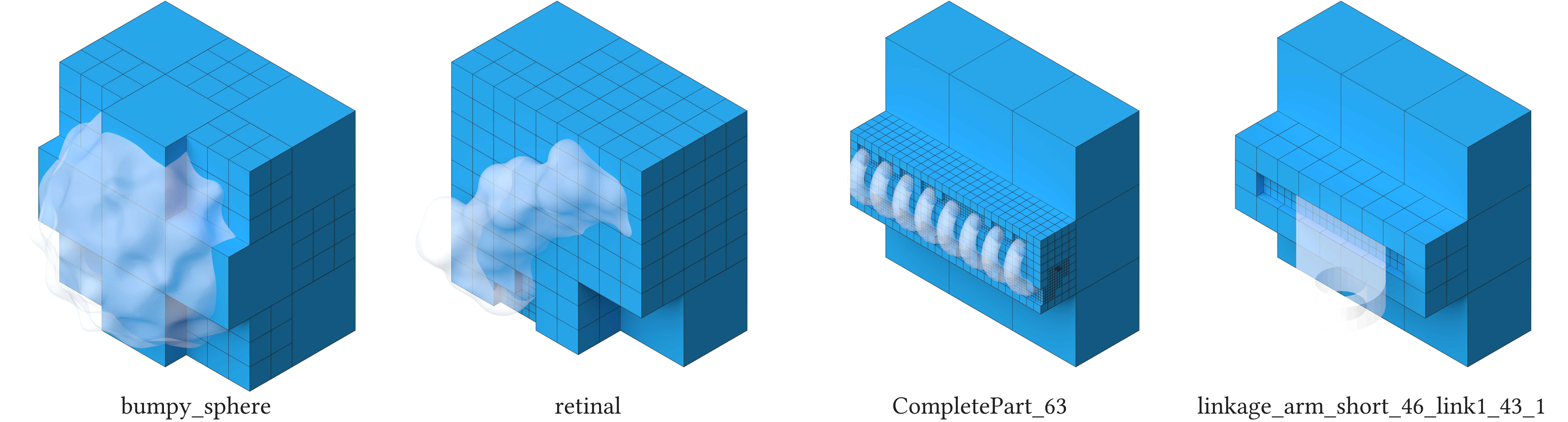}
\vspace{-8mm}
\caption{Four representative cases (selected from 202 models in \cite{gao2019feature}) where previous 3-refinement methods \cite{schneiders1996refining, ito2009octree, elsheikh2014consistent} outperform the proposed vertex-based or edge-based approaches in total hex element count. For geometries with mainly flat surfaces (e.g., planar faces or L-shaped sweeps), concave corners can be resolved by subdividing only a few grid cells. However, the concave-corner templates in the proposed methods generate more hex elements than that of 27 in full subdivision, giving previous methods an advantage in these specific scenarios.}
\label{fig:failCases}
\end{figure}

Figure \ref{fig:resultComparison} presents a comparative analysis among prior 3-refinement \cite{schneiders1996refining, ito2009octree, elsheikh2014consistent}, vertex-based, and edge-based methods. The relative growth metric, defined as the ratio of final hex elements to initial grid elements (with values closer to 1 indicating superior performance), is computed for each model in the dataset of \cite{gao2019feature}. The computation time of vertex-based and edge-based methods versus initial grid cell count is also exhibited. Table \ref{tab:resultComparison} quantifies the pairwise comparisons, enumerating instances where the proposed methods achieve lower (win), equal (tie), or higher (loss) growth ratios compared to each baseline.

Three patterns emerge. First, previous methods frequently refine the initial grid by an order of magnitude or more to enforce conformity, imposing substantial computational overhead. The vertex-based and edge-based methods constrain this expansion to a factor of five or less, achieving reductions ranging from several-fold to over a hundred-fold compared to prior approaches. Second, both proposed methods demonstrate comparable performance, substantially outperforming prior methods across most models. As shown in the upper-right panel of Figure \ref{fig:resultComparison}, computation time scales linearly with initial grid cell count for both methods, indicating favorable algorithmic complexity. The edge-based variant, due to its greedy search, requires approximately five times longer than the vertex-based approach given the same initial grid. In reward, the edge-based variant consistently yields marginally fewer elements, producing a lower count in 92\% models while matching the vertex-based count in the remaining 8\% models. Third, although prior methods occasionally achieve comparable or slightly better results, these cases are restricted to geometries with mainly flat surfaces (e.g., planar faces or L-shaped sweeps). As illustrated in Figure \ref{fig:failCases}, such configurations require minimal subdivision to resolve concave corners. The marginal advantage arises because vertex- or edge-based concave corner templates generate more than 27 elements per cell, exceeding the cost of full subdivision, thereby allowing the brute-force uniform refinement of previous methods to prevail incidentally. Notably, as the refinement ratio increases, the proposed methods reduce progressively more hex elements over prior methods, since higher ratios typically correspond to more complex adaptive tree transitions where sophisticated templates can fully exploit their design advantages.

\begin{figure}
\centering
\includegraphics[width=\linewidth]{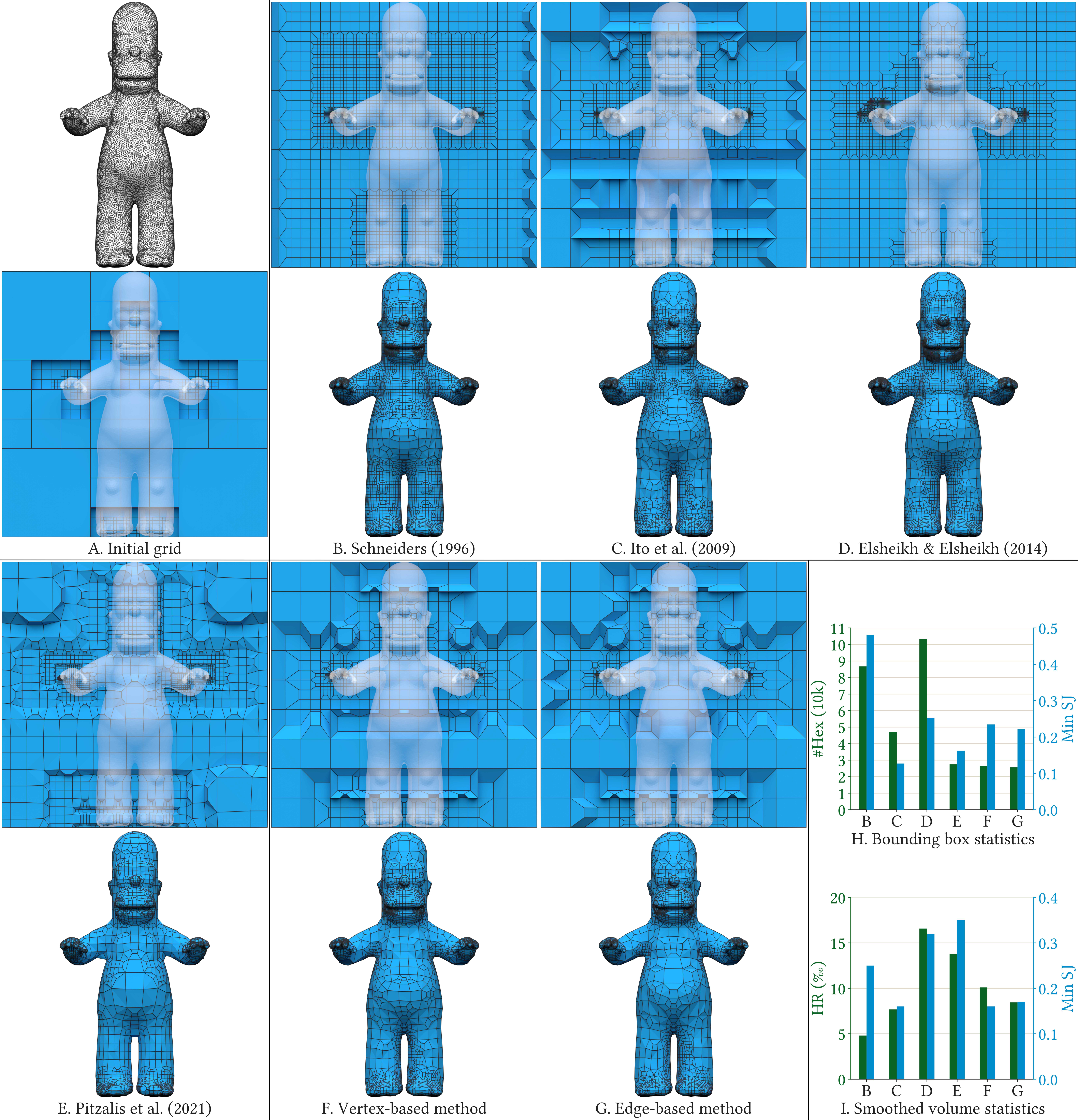}
\vspace{-8mm}
\caption{Adaptive refinement to conformal hex meshes. (A) Initial adaptively refined grid. (B-D) Previous 3-refinement methods produce dense grids. (E) The element-optimal 2-refinement method achieves significant reduction in element count. (F, G) With hex element count set slightly lower than (E), both vertex- and edge-based templates yield grids substantially coarser than (B-D). (H) Hex element count and min SJ in bounding-box hex meshing: (B-D), (F), and (G) are derived from the same initial grid (A) using identical shape diameter function parameter that yields (F) and (G) with slightly fewer hex elements than (E). (F) and (G) also have several-fold smaller hex element count than (B-D). (I) Hausdorff ratio and min SJ of smoothed volume meshes: (E) exhibits the highest min SJ, while (F, G) have smaller Hausdorff ratio errors.}
\label{fig:teaser}
\end{figure}

Figure \ref{fig:teaser} presents an overall comparison between the two proposed methods, three previous 3-refinement methods \cite{schneiders1996refining,ito2009octree,elsheikh2014consistent}, and the state-of-the-art 2-refinement method \cite{pitzalis2021generalized} applied to the identical input geometry. All 3-refinement methods operate on the same initial grid using the same shape diameter function parameter that ensure the bounding box hex element counts of two proposed methods to be marginally smaller than that of \cite{pitzalis2021generalized}. Against previous 3-refinement methods, two proposed methods introduce the fewest additional hex elements while maintaining similar template min SJ values. Moreover, they achieve better geometry fitting with slightly lower Hausdorff ratios than \cite{pitzalis2021generalized}.

\begin{figure}
\centering
\includegraphics[width=\linewidth]{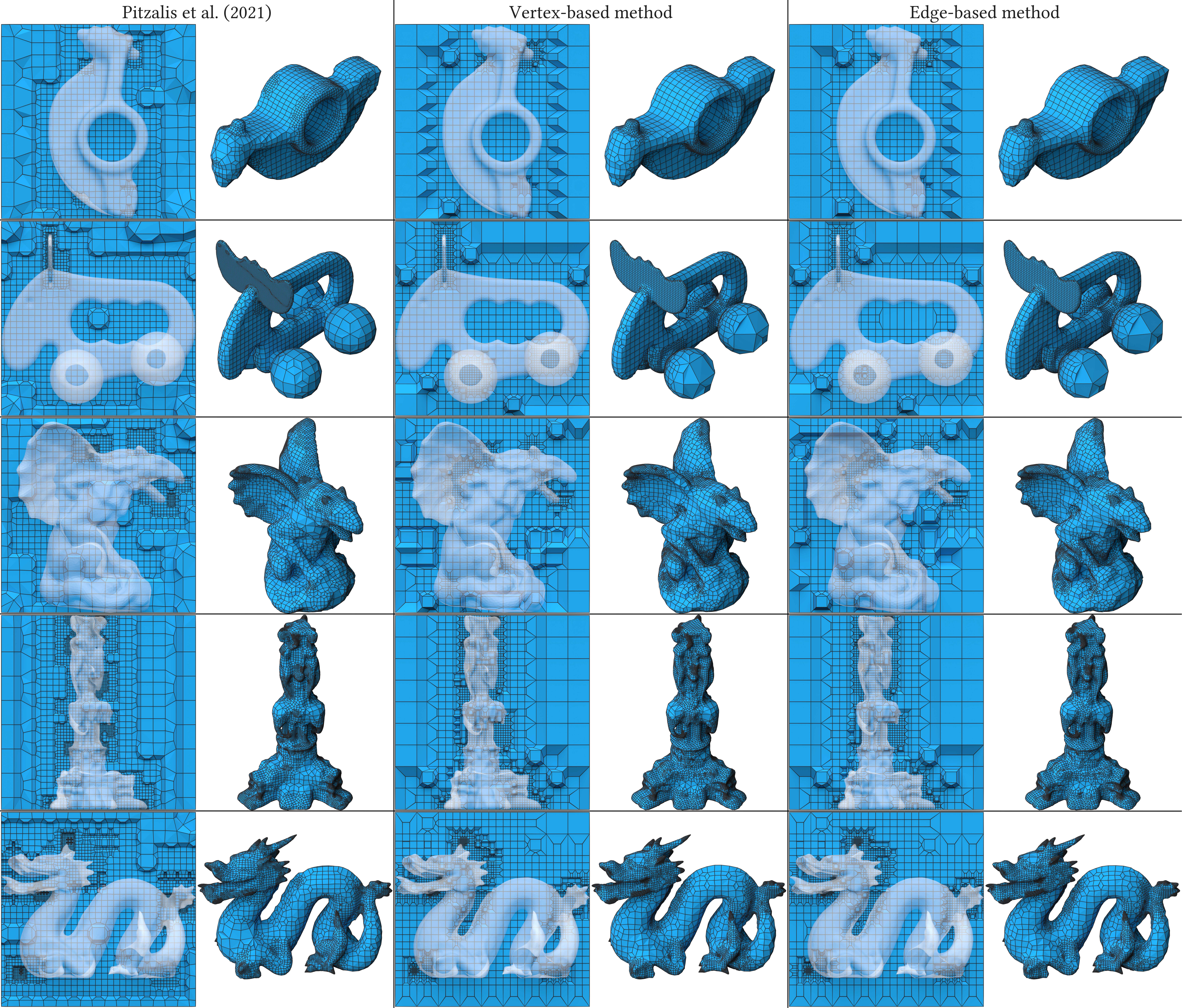}
\vspace{-8mm}
\caption{Evaluation on five randomly selected models (rocker, elk, gargoyle, thai\_statue and dragonstand2) from the benchmark dataset \cite{gao2019feature}. For each model, results are shown from three methods: \cite{pitzalis2021generalized}, vertex-based method, and edge-based method. Each method is visualized through a cross-section of its bounding box hex mesh and the corresponding smoothed volume mesh.}
\label{fig:compareWith2Refinement}
\end{figure}

\begin{table}
\centering
\caption{Statistical summary of Figure \ref{fig:compareWith2Refinement}. The table reports four metrics: template hex element count, template min SJ, smoothed volume mesh Hausdorff ratio, and smoothed volume mesh min SJ. The hex element counts for the proposed methods are adjusted via tuning the shape diameter function threshold to be slightly smaller than \cite{pitzalis2021generalized}. Each table cell displays \cite{pitzalis2021generalized} result (left), vertex-based result (middle) and edge-based result (right); best values are bolded. \cite{pitzalis2021generalized} shows advantage in min SJ, whereas the proposed two methods show advantage in HR.}
\label{tab:compareWith2Refinement}
\begin{tabular}{l|l|l|l|l}
& Template \#hex & Template min SJ\(\uparrow\) & Volume HR\(\downarrow\) & Volume min SJ\(\uparrow\) \\\hline
rocker & 15,584/15,126/14,992 & 0.095/\textbf{0.23}/0.22 & 1.1/1.0/\textbf{0.97} & \textbf{0.45}/0.20/0.20 \\\hline
elk & 30,788/30,767/28,716 & 0.22/\textbf{0.23}/0.22 & \textbf{1.6}/2.0/1.9 & \textbf{0.45}/0.22/0.23 \\\hline
gargoyle & 40,475/40,048/38,130 & 0.094/\textbf{0.23}/0.22 & 1.9/1.6/\textbf{1.5} & \textbf{0.39}/0.16/0.24 \\\hline
thai\_statue & 89,877/88,475/82,714 & 0.092/\textbf{0.23}/0.22 & 0.94/0.87/\textbf{0.80} & \textbf{0.26}/0.22/0.22 \\\hline
dragonstand2 & 114,483/114,253/110,440 & 0.093/\textbf{0.23}/0.22 & 1.0/0.98/\textbf{0.95} & 0.11/0.10/\textbf{0.13}
\end{tabular}
\end{table}

Next, a comparative evaluation is made between vertex- and edge-based methods and the state-of-the-art 2-refinement approach \cite{pitzalis2021generalized}. Figure \ref{fig:compareWith2Refinement} displays the results for five randomly selected test cases, with statistics provided in Table \ref{tab:compareWith2Refinement}. The reported metrics include bounding box hex mesh element count, bounding box hex mesh min SJ, smoothed volume mesh Hausdorff ratio relative to the input geometry, and smoothed volume mesh min SJ.

The results demonstrate that two proposed methods consistently outperform the baseline across most quality measures while using fewer hex elements. Specifically, they produce template meshes with stable, superior min SJ values (0.23 and 0.22, respectively), whereas the baseline exhibits lower values and considerable variability. This stability arises from the direct manipulation of the primal mesh, which yields an enumerable set of element configurations. In contrast, the dual mesh transformation employed by \cite{pitzalis2021generalized} introduces greater geometric uncertainty and quality fluctuations.

Additionally, the proposed methods achieve lower Hausdorff ratios in the majority of cases, indicating enhanced adaptivity to geometric features. Although \cite{pitzalis2021generalized} employs a weakly-balanced condition that saves elements more aggressively than moderately-balanced condition, its reliance on the generalized pairing condition for 2-refinement makes it less efficient, potentially over-refining low-interest regions while under-sampling critical features, thereby increasing the Hausdorff ratio. The only exception occurs around the four spheres in the elk model, where two proposed methods yield higher Hausdorff ratios. This stems from the initial grid resolution being too coarse to trigger one level deeper subdivision at the spheres. This limitation reflects the inherently steeper density transitions of 3-refinement relative to 2-refinement. Regarding smoothed volume mesh quality, the min SJ values of the two proposed methods are lower than those of \cite{pitzalis2021generalized}. This is attributed to the steeper density transitions in 3-refinement, which generate more irregular vertex and edge valence and consequently degrade mesh quality. Specifically, it is observed that poor-quality elements predominantly concentrate near boundaries, where highly irregular local vertex and edge valence constrains the achievable element quality. However, statistical analysis across these five models reveals no significant distinction in average mesh quality. The average scaled Jacobian values after optimization are 0.87, 0.81, 0.90, 0.74, and 0.78, respectively; these averages are identical across all three methods for each model.

To guide the choice between these approaches, it is essential to weigh the trade-offs among mesh quality, geometry fitting, and theoretical bound. The proposed vertex- and edge-based methods are preferable when geometry fitting and theoretical bound are important. In terms of geometry fitting, the 3-refinement templates refine more adaptively, characterized by steeper density transitions, thereby capturing small features more effectively. Regarding theoretical bound, this approach is superior because all hex element faces are planar, and the number of element types is significantly more concise than the dual hex elements generated by 2-refinement \cite{tong2025mchex}. Conversely, the 2-refinement approach excels in mesh quality; due to the more regular vertex and edge valences it produces, it achieves higher mesh quality within the same optimization time budget.

\section{Conclusion and Future Work}
\label{sec:conclusionAndFutureWork}

This paper introduces an element-efficient 3-refinement algorithm for hanging-node removal that significantly reduces hex element counts compared to existing 3-refinement methods. The key innovation is a recursive local refinement strategy that decomposes grid cells exhibiting complex vertex/edge refinement patterns into simpler cells. Cells matching a set of fundamental templates are replaced via direct template substitution until the entire grid is processed. This method require only a moderately-balanced condition, restricting edge refinement to \(1\rightarrow3\) transitions and face refinement to \(1\times1\rightarrow3\times3\) patterns, substantially more relaxed than prior 3-refinement constraints \cite{schneiders1996refining,ito2009octree,elsheikh2014consistent}. Evaluations on 202 models in the \cite{gao2019feature} dataset demonstrate that, given same initial grids, the proposed method reduces final hex element counts by several-fold to several-hundred-fold. Moreover, while using marginally fewer elements than the state-of-the-art 2-refinement method \cite{pitzalis2021generalized}, the proposed approach achieves more accurate geometric fitting with lower Hausdorff ratios. Crucially, all generated hex elements have coplanar quad faces, which is important in some application scenarios.

Several limitations and promising directions for future research are identified. First, empirical evidence demonstrates that the edge-based greedy algorithm yields varying reduction results depending on the traversal order of unrefined edges. Although these variations are typically minor, they underscore the suboptimal nature of greedy approaches and may compromise code reproducibility. Future work should explore advanced element reduction techniques that achieve more aggressive simplification within acceptable computational time while guaranteeing order-independent outcomes. Second, the current pipeline does not account for mesh connectivity metrics, specifically vertex and edge valence. A promising avenue for future research involves mitigating valence irregularity through grid cell reorientation while preserving six cell face patterns. This would mitigate high valence configurations by reducing the number of adjacent hex elements at irregular vertices and edges, thereby enhancing overall mesh regularity. Third, Artificial Intelligence (AI) is beginning to reduce the manual effort in the meshing pipeline \cite{tong2023srl,owen2025survey,yu2025dl,yu2026ddpm}. A promising future direction involves utilizing reinforcement learning to explore more optimal hex transition templates, aiming to generate meshes with fewer elements and more regular valence configurations.

\bibliography{mybibfile}

\end{document}